\def\preprint{TUW-99/23}        

\def\archive {hep-th/9910067}           

\def\Title{ \Huge   Killing gauge for the 0-brane on
\\[9pt]			$AdS_2 \times S^2$ coset superspace         
}
\long\def\Abstract{
	How to gauge fix $\k$-symmetry for the super 0-brane action 
	on $AdS_2 \times S^2$ in Killing gauge properly is discussed
	in order to find the superconformal mechanics which describes
	super 0-brane probes moving on $AdS_2 \times S^2$. The dependence on
	the coordinate frame for the proper Killing gauge is considered
	and the subtleties of gauge-fixing $\k$-symmetry in Killing gauge
	are analysed explicitly. It is found that the Killing gauge works 
	indeed without the imcompatibility
	if the magnetic charge of the super 0-brane is nonzero.
}
\def\fnote#1#2{\begingroup\def\thefootnote{#1}\footnote{#2}
                \addtocounter{footnote}{-1}\endgroup}   
\let\3=\ss	\catcode`\"=\active \let"=\"

\documentclass[12pt]{article} 

\newcommand{\EQ}{\begin{equation}}
\newcommand{\EN}{\end{equation}}
\newcommand{\bea}{\begin{eqnarray}} 
\newcommand{\ena}{\end{eqnarray}}
\newcommand{\vs}[1]{\vspace{#1 mm}}

\renewcommand{\a}{\alpha}
\renewcommand{\b}{\beta}
\renewcommand{\c}{\gamma}
\renewcommand{\d}{\delta}
\newcommand{\e}{\epsilon}
\newcommand{\C}{\Gamma}
\def\bbox{{\,\lower0.9pt\vbox{\hrule \hbox{\vrule height 0.2 cm
\hskip 0.2 cm \vrule height 0.2 cm}\hrule}\,}}
\newcommand{\dsl}{\pa \kern-0.5em /}

\newcommand{\pa}{\partial}
\renewcommand{\t}{\theta}
\renewcommand{\k}{\kappa}

\def\apr{{a^\prime}}
\def\bpr{{b^\prime}}
\def\cpr{{c^\prime}}

\def\alpr{{\alpha^\prime}}
\def\bepr{{\beta^\prime}}
\def\gapr{{\gamma^\prime}}

\newcommand{\nn}{\nonumber\\}
\newcommand{\p}[1]{(\ref{#1})}

\begin{document}

\topmargin 0pt
\oddsidemargin 0mm
\renewcommand{\thefootnote}{\fnsymbol{footnote}}
\begin{titlepage}

\setcounter{page}{0}
\begin{flushright}
\archive   \vskip -2pt \hfill\preprint 			%% hep-th/9910xxx
\end{flushright}

\vs{10}
\begin{center}

{\Large\bf \Title	}

\vs{15}

{\large }

\vs{10}

        Maximilian KREUZER\fnote{*}{e-mail: kreuzer@hep.itp.tuwien.ac.at}
	and Jian-Ge ZHOU\fnote{\#}{e-mail: jgzhou@hep.itp.tuwien.ac.at}
\\[3mm]
        Institut f"ur Theoretische Physik, Technische Universit"at Wien\\
        Wiedner Hauptstra\3e 8--10, A-1040 Wien, AUSTRIA

\end{center}

\vs{15}
\centerline{{\bf{Abstract}}}
\vs{5}
\Abstract
\vs{5}
\begin{flushleft}
\end{flushleft}

\end{titlepage}
\newpage

\renewcommand{\thefootnote}{\arabic{footnote}}
\setcounter{footnote}{0}

Recently, there has been much interest in the AdS/CFT correspondence
~\cite{MGW}, which relates string theory on AdS$_{p+2} \times
S^{D-P-2}$ to extended superconformal theories in $p+1$ dimensions.
In view of the AdS/CFT conjecture, it is important to understand
the formulation of superstrings and super p-branes on these curved
spaces. In~\cite{MT1}, the type IIB Green-Schwarz (GS) superstring 
action was constructed in AdS$_5 \times S^5$ background in terms of 
supercoset formalism. This action possesses global $SU(2,2\mid 4)$
super-invariance,  has $\k$-symmetry and 2D reparametrization
invariance as its local symmetries, and reduces to the conventional
type IIB GS superstring action in the flat background
limit. The other related construction for GS superstring, super 
D3-brane, D1-brane on AdS$_5 \times S^5$, and super M-branes on  
AdS$_4 \times S^7$
and AdS$_7 \times S^4$ have been discussed in~\cite{KRR}-\cite{O1}.
The GS superstring and super p-brane actions on 
AdS$_3 \times S^3$ and AdS$_2 \times S^2$ have been constructed
in~\cite{P3} and~\cite{JGZ}\footnote{A slightly different construction
for superstring on AdS$_2 \times S^2$ and AdS$_3 \times S^3$ was
discussed in~\cite{BB} and ~\cite{B1}. }. The gauge-fixing of
$\k$-symmetry was carried out in 
Killing gauge~\cite{K1} or supersolvable algebra approach~\cite{DF,
P1}. However, the $\k$-symmetry gauge-fixing
and quantization seem
still to pose some difficulties~\cite{PST, RR, P2}. In~\cite{PST}, it
was argued that there is an imcompatibility between Killing gauge and
the static vacuum solution for super p-brane actions on  AdS$_{p+2} \times
S^{D-P-2}$ superbackgrounds\footnote{As the superstring $\k$-symmetry 
projector is different from the reduced D3-brane projector in the
static vacuum solution, such an imcompatibility does not exist for 
GS superstring.}. Since the D3-brane action on 
AdS$_5 \times S^5$ is very complicated~\cite{MT2}, 
while the super 0-brane action on AdS$_2 \times S^2$ was constructed in 
supercoset formalism~\cite{JGZ} only recently,
% and only recently the super 0-brane action on AdS$_2 \times S^2$ 
% was constructed in supercoset formalism~\cite{JGZ}, therefore, 
it is quite interesting to see how the imcompatibility
mentioned in~\cite{PST}
appears explicitly and whether it is possible to simplify the super
0-brane action on AdS$_2 \times S^2$ in Killing gauge.

On the other hand, the radial motion of
a superparticle with zero angular momentum near the horizon
of an extreme Reissner-Nordstr\"om black hole (AdS$_2 \times S^2$)
is found to be described by an $Osp(1 \mid 2)$-invariant
superconformal mechanics~\cite{CDK},
and it was argued in ~\cite{CDK} that the full superparticle
dynamics should be invariant under the larger $SU(1,1 \mid 2)$
superconformal group because this is the superisometry group of 
AdS$_2 \times S^2$. This full dynamics describes not only the radial
motion of the superparticle, but also its motion on $S^2$. In ~\cite{AIB},
the authors tried to construct a $SU(1,1 \mid 2)$-invariant action from
the worldline superfield formalism, but, due to technical
difficulties, the explicit $SU(1,1 \mid 2)$-invariant action has
not been obtained. Then it is interesting to see how the explicit
action, which describes the super 0-brane dynamics on AdS$_2 \times
S^2$ with underlying $SU(1,1 \mid 2)$ invariance\footnote{By
``underlying $SU(1,1 \mid 2)$ invariance'' we mean that upon gauge 
fixing $\k$-symmetry, its superconformal transformations are
non-linearly realized on the remaining fields~\cite{DS, BLPS}.}, 
is constructed
in the bosonic and fermionic coordinates of the super 0-brane.

The purpose of this paper is that we explore the possibility
of how to simplify the super 0-brane action on AdS$_2 \times S^2$ 
in Killing gauge and avoid the imcompatibility in~\cite{PST} in order
to find the super 0-brane action on AdS$_2 \times S^2$ with
underlying $SU(1,1 \mid 2)$ invariance in terms of the 
bosonic and fermionic coordinates of the super 0-brane.
To achieve this goal, we exploit the super 0-brane action on AdS$_2 \times
S^2$ built out of the Cartan 1-forms $L^{a}$, $L^{\apr}$ and
$L^{I}$~\cite{JGZ}, which has global 
$SU(1,1 \mid 2)$-invariance, and is invariant under local
$\k$-symmetry and one-dimensional reparametrization symmetry.
The crucial feature of the super 0-brane action on AdS$_2 \times
S^2$ is that it contains two free parameters A and B, which
can be interpreted as the electric and magnetic charges of the
super 0-brane. To gauge fix $\k$-symmetry 
%% of the super 0-brane action on AdS$_2 \times S^2$ 
in Killing gauge~\cite{K1} and to avoid
the imcompatibility~\cite{PST}, we choose the magnetic charge (B) of the
super 0-brane to be nonzero. The 0-brane projector on
AdS$_2 \times S^2$  is given by ${\cal P}_{\pm}=
\frac{1}{2}({\d}^{IJ} \pm {\c}^{0}\epsilon^{IJ})$, where the signs
depend on the choice of the coordinate frames. First we consider the
Killing gauge in the coordinate frame \p{f1} whose Killing
horizon is at $r=\infty$~\cite{CDK}. We find that the proper
Killing gauge is ${\cal P}_{-}\Theta = \Theta_{-} = 0$, which
makes  $({\cal M}^2_{fix})D{\Theta}_{+} = 0$, and $(D\Theta_{+})^{I} 
= (\frac{\Lambda}{r}d\theta_{+})^{I}$, thus we can simplify the
Cartan 1-forms $L^{a}$, $L^{\apr}$ and $L^{I}$. If we work in AdS
coordinates, the situation is reversed, instead of $\Theta_{-}$
we have to put $\Theta_{+} = 0$ as Killing gauge to simplify
the Cartan 1-forms. The rule is that for a  metric $g_{00} \sim
r^{l}$, we pick $\Theta_{-} = 0$ if $l < 0$, but 
we have to choose  $\Theta_{+} = 0$ if $l > 0$.
With the simplified expression for the Cartan 1-forms, the
$\k$-symmetry gauge fixed super 0-brane action on AdS$_2 \times
S^2$ is obtained, which can be considered as the supersymmetric
generalization of the action in~\cite{CDK} with underlying
$SU(1,1 \mid 2)$ invariance. To get 
$\k$-symmetry gauge fixed super 0-brane action, we have taken
the magnetic charge of the super 0-brane to be nonzero.
To see the imcompatibility explicitly, we choose the parameter
$B = 0$ and $A > 0$, then the classical static vacuum solution
exists. For this static solution the gauge fixed super 0-brane
action vanishes, and the $\k$-symmetry transformation is reduced
to $\d_{\k}\Theta_{+}^{I} = \k_{+}^{I}$, which means that the gauge
fixing %% for $\k$-symmetry 
should be $\Theta_{+}^{I} = 0$ instead
of $\Theta_{-}^{I} = 0$ in \p{gf}, that is, the usual Killing
gauge is imcompatiable with the classical static vacuum solution.
Since in AdS coordinates, the proper gauge fixing for 
$\k$-symmetry is  $\Theta_{+} = 0$, naively it seems the
imcompatibility could be resolved. However, we find that in 
AdS coordinates, to make the action \p{ab} vanish in the
static solution \p{ss}, we have to choose $A = -m$. Then the
$\k$-symmetry transformation is changed into 
$\d_{\k}\Theta_{-}^{I} = \k_{-}^{I}$, which indicates that
the gauge fixing %% for $\k$-symmetry 
should be $\Theta_{-} = 0$
instead of $\Theta_{+}^{I} = 0$. Thus the imcompatibility
between Killing gauge and static solution cannot be smoothed
out by a change of the coordinate frame. But when the
magnetic charge B is nonzero, this imcompatibility can be
avoided and the simplification of the super 0-brane action
on AdS$_2 \times S^2$ in Killing gauge works, since
the static vacuum solution does not exist for $B \not= 0$.
Finally the invertibility of the fermionic kinetic operator
is discussed, and we find that if we choose gauge fixing
properly for the coordinate e, the Killing gauge fixing is
acceptable.

Now let us consider the super 0-brane action on Ad$S_2 \times S^2$
background  described in terms of the supercoset formalism\footnote{We use
the conventions in ref.~\cite{JGZ}}~\cite{JGZ}
\bea
I_{0-brane}&= &-m\int{dt \sqrt {-({L^a_0}{L^a_0} +
{L^{\apr}_0}{L^{\apr}_0})}}\nn
&& 
+ \int\limits_{{\cal M}_2}\{A\e_{IJ}{\bar L}^I \wedge L^{J}
+ B\e_{IJ}{\bar L}^{I}\C_{5} \wedge L^{J}\nn
&& 
- \frac{A}{2}\e_{ab}L^a \wedge L^b - 
\frac{B}{2}\e_{\apr\bpr}L^{\apr}\wedge L^{\bpr}\}
\label{ba}
\ena
where the parameters m, A and B are the mass, electric and magnetic
charges of the super 0-brane respectively. The $\k$-symmetry 
transformation is defined by~\cite{JGZ}\footnote{The interesting
property of $\k$-symmetry of GS actions in coset superspaces was
discussed in~\cite{MC}.} 
\bea
\d_{\k}x^a = 0, \;\;
\d_{\k}x^{\apr} = 0,\nn
\d_{\k}\t^I = [(1+\C)\k)]^I
\label{ks}
\ena
with
\bea
\C  = {\frac{(A - B\c\otimes\gapr)(L^{a}_{0}\c^{a} +
L^{\apr}_{0}\c\otimes\c^{\apr})}{\sqrt{-(A^{2}+B^{2})
(L^{a}_{0}L^{a}_{0} + L^{\apr}_{0}L^{\apr}_{0})}}}{\cal E}
\label{pro}
\ena
\bea
{\cal E}= \pmatrix{
0  & -1 \cr 1 & 0 \cr }\label{epsilon}
\ena
\EQ
m = \sqrt {A^{2}+B^{2}}
\label{ma}
\EN
where the expression for $\k$-symmetry includes the parameters A and B,
and $m = \sqrt {A^{2}+B^{2}}$ occurs as a consequence of $\k$-symmetry
of the super 0-brane action.

The invariant 1-forms $L^{I}= L^{I}_{s=1}, L^{\hat a}= L^{\hat
a}_{s=1}$ are given by~\cite{KRR}
\bea
L^{I}_{s}&=&\left(\left (\sinh\left(s{\cal M}\right) \over {{\cal
M}}\right) D\Theta \right)^{I}\nn
L_s^{\hat a }&=&e^{\hat a }_{\hat m} (x) dx^{\hat m} + 4\bar
\Theta^I\Gamma^{\hat a}
\left({\sinh^2  \left({s\cal M}/2\right) \over {\cal M}^2} D\Theta
\right)^I
\label{form}
\ena
where $x^{\hat m}$ and $\Theta^I$ are the bosonic and fermionic
super 0-brane coordinates and for the $SU(1,1 \mid 2)$ superalgebra we have
\bea
({\cal M}^2)^{ IL}&=&-\epsilon^ {IJ}\c (\c_{ a} \Theta^{J} \bar
\Theta^L \c^{ a} + \c\otimes\gamma_{ a'}
\Theta^{J} \bar \Theta^L \c\otimes\gamma^{ a'} ) \nn
&+& {1\over 2}
\epsilon^{KL} (-\gamma_{ab} \Theta^I \bar \Theta^K \gamma^{ab}
+\gamma_{a'b'} \Theta^I \bar \Theta^K \gamma^{a'b'})\c,\nn
(D\Theta)^I&=&[d + {\frac{1}{4}}(\omega^{ab}\c_{ab} +
\omega^{\apr\bpr}\c_{\apr\bpr})]\Theta^I\nn
&+& {1\over 2}
\epsilon^{IJ}(e^{a}\c_{a}\c - e^{\apr}\c_{\apr})\Theta^J
\label{md}
\ena
where the Dirac matrices are split in a `2+2' way.

Before gauge fixing the $\k$-symmetry for the super 0-brane action
on Ad$S_2 \times S^2$, we choose the coordinates ~\cite{CDK}
\bea
ds^2&=&-{\left(\frac{2M}{r}\right)}^{4}{d\tau}^{2} + {\left
(\frac{2M}{r}\right)}^{2}dr^{2}
+ M^{2}\left(d{\chi}^{2} + {\sin^{2}{\chi}}d{\phi}^2\right)
\label{f1}
\ena
where the Killing horizon in these coordinates is at $r=\infty$. In
the following discussion, we put $M=1$ for simplicity, and
finally we recover M by dimension analysis.

Since the superstring $\k$-symmetry projector differs from the
D3-brane projector in Ad$S_5 \times S^5$, the Killing spinor
gauge works for the GS superstring action on Ad$S_5 \times S^5$
~\cite{KR1}, and a similar conclusion holds for the GS superstring action on
Ad$S_2 \times S^2$. In ~\cite{PST}, it is argued that there is
imcompatibility between Killing spinor gauge for the D3-brane action
on Ad$S_5 \times S^5$ and the static vacuum solution of the D3-brane
equation of motion. Similarly, this imcompatibility also exists for the
1-brane action on Ad$S_3 \times S^3$ and for the 0-brane action on Ad$S_2
\times S^2$. To make use of the Killing spinor gauge  and avoid
the imcompatibility, we have to
choose the magnetic charge (B) of super 0-brane to be nonzero (the
reason will become clear below). If we define ${\cal
P}_{\pm}=\frac{1}{2}({\d}^{IJ} \pm {\c}^{0}\epsilon^{IJ})$, 
${\cal P}_{\pm}\Theta = {\Theta}_{\pm}$, one may wonder which component,
${\Theta}_{+}$ or ${\Theta}_{-}$,
is put to zero for gauge fixing $\k$-symmetry in the coordinate frame \p{f1}. 
Even though the projector is indicated by the full 0-brane Killing
spinor~\cite{K1, KR1}, as we will see below, it is coordinate frame dependent.

In the coordinate frame \p{f1}, the proper gauge is 
\bea
{\cal P}_{-}\Theta = \Theta_{-} = 0, \; \; 
\epsilon^{IJ}\Theta_{+}^{J} 
= -\c^{0}\Theta_{+}^{I}.
\label{gf}
\ena
We will show that in this gauge we have $({\cal
M}^2_{fix})D{\Theta}_{+} = 0$, which means that all terms of the
$({\cal M}^{2n}_{fix})D{\Theta}_{+}$ for $n > 0$ vanish. Since the
coordinate frame \p{f1} possesses the property $\omega^{01} + e^{0} = 0$,
one has\footnote{If we choose ${\Theta}_{+}$ to be zero instead of
${\Theta}_{-}$, we cannot obtain \p{d1} in the coordinate frame \p{f1}.}
\bea
\epsilon^{IJ}(D\Theta_{+})^{J}& = & -\c^{0}(D\Theta_{+})^{I},
\label{d1}
\ena
Here we should emphasize that, to get \p{d1},  $\omega^{01} + e^{0} = 0$
plays a crucial role. With \p{gf} and \p{d1}, one can easily
show that if $\{\c^{0}, U\} = 0$\footnote{Unlike GS superstring on
AdS$_5 \times S^5$, there only if $[ \Gamma^{0123}, U ] = 0$, \p{zero}
holds.}, one has
\bea
\bar \Theta_{+}^{I} U D{\Theta_{+}}^{I} = 0.
\label{zero}
\ena
By exploiting \p{gf}, \p{d1} and \p{zero}, one gets
\bea
({\cal M}^{2}_{fix})^{IL}D{\Theta}_{+}^{L}& = & -\epsilon^{IJ}\c
\left(\c_{a}\Theta^{J}_{+}\bar \Theta^{L}_{+}\c^{a} + \c\otimes\c_{\apr}
\Theta^{J}_{+}\bar\Theta^{L}_{+}\c\otimes\c^{\apr}\right)D{\Theta}_{+}^{L} \nn
&+& {\frac{1}{2}}
\epsilon^{KL}\left(-\c_{ab}\Theta^{I}_{+}\bar
\Theta^{K}_{+}\c^{ab} + \c_{\apr\bpr}\Theta^{I}_{+}\bar
\Theta^{K}_{+}\c^{\apr\bpr}\right){\c}D{\Theta}_{+}^{L} = 0.
\ena
Then the 1-forms are simplified as
\bea
\left(L^{I}_{s}\right)_{+} = sD\Theta^{I}_{+}, \;\; 
\left(L^{I}_{s}\right)_{-} = 0 \nn
L^{0}_{s} = \left(\frac {2}{r}\right)^{2}d\tau +
s^{2}\bar\Theta^{I}_{+}\c^{0}D\Theta^{I}_{+} \nn
L^{r}_{s} =\frac {2}{r} dr, \;\;
L^{\chi}_{s} = d\chi, \;\;
L^{\phi}_{s} = \sin\chi d\phi.
\label{nf}
\ena
Since we are interested in a `2 + 2' spliting, we need work out
$D\Theta_{+}$ explicitly in the coordinate frame \p{f1}, which
can be expressed as
\bea
D\Theta_{+}^{I}& = & \left [ d + \frac {1}{4}(\omega^{ab}\c_{ab}
+ \omega^{\apr\bpr}\c_{\apr\bpr})\right]\Theta_{+}^{I} \nn
&+& \frac{1}{2}
\epsilon^{IJ}(e^{a}\c_{a}\c - e^{\apr}\c_{\apr})\Theta_{+}^{J} \nn
&=& \frac {\Lambda}{r}d(r\Lambda^{-1}\Theta_{+})^{I}\label{d2},
\ena
where we have used the fact that
\bea
 \frac {1}{4}\d^{IJ}\omega^{\apr\bpr}\c_{\apr\bpr} -  \frac{1}{2}
\epsilon^{IJ}e^{\apr}\c_{\apr} = (\Lambda d\Lambda^{-1})^{IJ}
\ena
with
\bea
\Lambda = e^{-{\chi}{\cal E}\c_{3}/2}e^{-{\phi}\c_{23}/2},
\label{rot}
\ena
where ${\cal E}$ is defined in \p{epsilon}, and in deriving \p{d2},
we have exploited the property $\omega^{01} + e^{0} = 0$. If we define
new variables
\bea
\theta^{I}_{+} = r(\Lambda^{-1}\Theta_{+})^{I}, \;\;
\Theta_{+}^{I} = \frac{1}{r}(\Lambda\theta_{+})^{I}
\label{nv}
\ena
then $D\Theta_{+}^{I}$ turns into 
\bea
(D\Theta_{+})^{I} = (\frac{\Lambda}{r}d\theta_{+})^{I}.
\label{nd}
\ena
With \p{nd}, the 1-forms can be further reduced to
\bea
\left(L^{I}_{s}\right)_{+} = \frac{s}{r}({\Lambda}d\theta_{+})^{I}, \;\; 
\left(L^{I}_{s}\right)_{-} = 0\,, \nn
L^{0}_{s} = \left(\frac {2M}{r}\right)^{2}\left(d\tau +
\frac{s^{2}}{4}\bar\theta^{I}_{+}\c^{0}d\theta^{I}_{+}\right), \nn
L^{r}_{s} = \frac{2M}{r}dr, \;\;
L^{\chi}_{s} = Md\chi, \;\;
L^{\phi}_{s} = M\sin\chi d\phi \,,
\label{of}
\ena
where we have exploited the explicit expression for $\Lambda$, the
relation $\c^{\apr} = {\cpr}^{-1}{\c^{\apr}}^{T}{\cpr}$~\cite{JGZ}, and the 
dependence on M has been recovered. We notice that the dependence of
the 1-forms on $\Lambda$ has been removed.

To get \p{of}, we have heavily exploited the equation  $\omega^{01} +
e^{0} = 0$, and the gauge fixing for $\k$-symmetry has to be taken as
$\Theta_{-} = 0$ in order to use \p{d1} and \p{d2} to get
${\cal M}^{2}_{fix}D{\Theta}_{+} = 0$. If we instead choose
${\Theta}_{+} = 0$ to gauge fix the  $\k$-symmetry, the simplification
cannot be carried out since \p{d1} and \p{d2} fail in the coordinate
frame \p{f1}. However, in AdS (spherical) coordinates\footnote{To
get \p{f2}, we need do the transformation $r \rightarrow
2Mr^{-\frac{1}{2}}$ from \p{f1}.}
\bea
ds^2 & = &-\left(\frac{r}{M}\right)^{2}{d\tau}^{2} + {\left
(\frac{M}{r}\right)}^{2}dr^{2}
+ M^{2}\left(d{\chi}^{2} + {\sin^{2}{\chi}}d{\phi}^2\right)
\label{f2}
\ena
%% in which 
the Killing horizon is %% put 
at $r = 0$ and %% In AdS coordinates \p{f2}, 
we have $\omega^{01} - e^{0} = 0$. From the above discussion
we know that the proper gauge fixing for $\k$-symmetry is 
\bea
{\cal P}_{+}\Theta = \Theta_{+} = 0, \; \; 
\epsilon^{IJ}\Theta_{-}^{J} 
= \c^{0}\Theta_{-}^{I},
\label{gf1}
\ena
which reverses the role of %% position between 
$\Theta_{+}$ and $\Theta_{-}$. 
Moveover, we have
\bea
\epsilon^{IJ}(D\Theta_{-})^{J} =  \c^{0}(D\Theta_{-})^{I}, \nn
({\cal M}^2_{fix})^{IL}(D{\Theta}_{-})^{L} = 0, \nn
(D\Theta_{-})^{I} = r^{\frac{1}{2}}({\Lambda}d\theta_{-})^{I}, \nn
\theta^{I}_{-} = r^{-\frac{1}{2}}(\Lambda^{-1}\Theta_{-})^{I}
\label{ads}
\ena
and the corresponding 1-forms in AdS coordinates are simplified as 
\bea
\left(L^{I}_{s}\right)_{-} = sr^{\frac{1}{2}}({\Lambda}d\theta_{-})^{I}, \;\; 
\left(L^{I}_{s}\right)_{+} = 0 \nn
L^{0}_{s} = \frac{r}{M}\left(d\tau +
s^{2}\bar\theta^{I}_{-}\c^{0}d\theta^{I}_{-}\right), \nn
L^{r}_{s} = \frac{M}{r}dr, \;\;
L^{\chi}_{s} = Md\chi, \;\;
L^{\phi}_{s} = M\sin\chi d\phi.
\label{adsf}
\ena
What we learned is that, for a given coordinate frame, the Killing gauge
fixing for $\k$-symmetry is unique: For $g_{00} \sim r^{l}$ 
 we have to put $\Theta_{-} = 0$ when $l< 0$, and we should
choose  $\Theta_{+} = 0$ when $l > 0$.

To get the superconformal mechanics for the super 0-brane on AdS$_2 \times
S^2$, we consider the coordiante frame \p{f1}. To represent the
WZ term in \p{ba} as an integral % over the one-dimensional space, 
we use the standard trick of rescaling $\Theta \rightarrow \Theta_{s}
\equiv s\Theta$, 
\bea
I_{WZ} = I_{WZ}(s = 1), \;\;
\pa_{s}I_{WZ}(s) = \int_{\pa{\cal M}_{2}}\pa_{s}H(s)
\ena
and
\bea
\pa_{s}H(s) = 2\left( A\epsilon^{IJ}\bar\Theta^{I}L^{J}_{s}
+ B\epsilon^{IJ}\bar\Theta^{I}\Gamma_{5}L^{J}_{s}\right),
\ena
where we have used the following equation~\cite{JGZ}:
\bea
\d{{\cal L}_{WZ}} = -2\left( A\epsilon^{IJ}\bar L^{I}\d\Theta^{J}
+ B\epsilon^{IJ}\bar L^{I}\Gamma_{5}\d\Theta^{J}\right)
\ena
Then we have
\bea
I_{WZ}(s = 1) = I_{WZ}(s = 0) + 2\int\limits_{0}^{1}ds\int dt\left(
A\epsilon^{IJ}\bar\Theta^{I}L^{J}_{s}
+ B\epsilon^{IJ}\bar\Theta^{I}\Gamma_{5}L^{J}_{s}\right)\label{wz}
\ena
In the coordinate frame \p{f1}, we get
\bea
I_{WZ} = \int dt \left
[ A\left(\frac{2M}{r}\right)^{2}\dot{\tau} + BM {\cos\chi}
\dot{\phi} + A\epsilon^{IJ}\bar\Theta^{I}_{+}D\Theta^{J}_{+}\right],
\ena
where we have
used \p{d1} and \p{zero} to show that $\epsilon^{IJ}\bar\Theta_{+}^{I}
\Gamma_{5}(D\Theta_{+})^{J} = - \bar\Theta_{+}^{I}
\Gamma_{5}\c^{0}(D\Theta_{+})^{I}$ vanishes. 
We note that the first two terms in the brackets come from $I_{WZ}(s = 0)$,
which vanishes in the case of the GS superstring action. Then the 
$\k$-symmetry gauge fixed super 0-brane action on AdS$_2 \times S^2$
is
\bea
I_{0-brane}&=& \int dt
\left\{-m\left(\frac{2M}{r}\right)^{2}\left[\left(\dot{\tau} +
\frac{1}{2}\bar\theta_{+}\c^{0}\dot{\theta}_{+}\right)^{2} -
\frac{r^{2}\dot{r}^{2}}{4M^{2}} - M^{2}\left(\frac{r}{2M}\right)^{4}
\left(\dot{\chi}^{2} +
{\sin^{2}\chi}\dot{\phi}^{2}\right)\right]^{\frac{1}{2}}\right\} \nn
&+& \int dt\left\{A\left(\frac{2M}{r}\right)^{2}\left(\dot{\tau} - \frac{1}{2}
\bar\theta_{+}\c^{0}\dot{\theta}_{+}\right) + BM {\cos\chi}
\dot{\phi}\right\},
\label{ob}
\ena
where $\theta_{+}$ denotes $\theta_{+}^1$. Eq.\p{ob} describes the 
dynamics of the super 0-brane in AdS$_2 \times
S^2$ background, which generalizes the action given in~\cite{CDK}. By
introducing the auxiliary coordinate e, the above action is rewritten
as
\bea
I_{0-brane}&=& \int dt
\left\{\frac{1}{2}e^{-1}\left[-\left(\frac{2M}{r}
\right)^{4}\left(\dot{\tau} +
\frac{1}{2}\bar\theta_{+}\c^{0}\dot{\theta}_{+}\right)^{2}
+ \frac{4M^{2}\dot{r}^{2}}{r^{2}} + M^{2}\left(\dot{\chi}^{2} +
{\sin^{2}\chi}\dot{\phi}^{2}\right)\right] - \frac{1}{2}em^{2}\right\} \nn
&+& \int dt\left\{A\left(\frac{2M}{r}\right)^{2}\left(\dot{\tau} - \frac{1}{2}
\bar\theta_{+}\c^{0}\dot{\theta}_{+}\right) + BM {\cos\chi}
\dot{\phi}\right\}
\label{eb}
\ena
Variating action \p{eb} with respect to the variable $\chi$, we
have $\ddot{\chi} \sim B{\sin\chi} \dot{\phi}$, which shows that
we can interpret B as the magnetic charge of the super 0-brane.

To get \p{ob}, we have assumed the magnetic charge of the
super 0-brane to be nonzero. When we choose the parameter $B = 0$
and $A > 0$, we have $A = m$ from \p{ma} and  there is a classical
static vacuum solution  of the super 0-brane equation of motion
following from \p{ob}~\cite{PST},
\bea
\tau = t, \;\;
r = constant, \;\;
\chi = constant, \;\;
\phi = constant, \;\;
\theta_{+} = 0.
\label{ss}
\ena
For this static solution, the action \p{ob} vanishes ($A > 0$,
$B = 0$), which is called the no-force condition, since there is
no potential which can push the super 0-brane probe to the 
boundary of Ad$S_{2}$. When $A > 0$, $B = 0$, the $\k$-symmetry
projector is reduced to  $\Gamma = -{\cal E}\c^{0}$ and \p{ks}
can be written as 
\bea
\d_{\k}\Theta^{I} = (\d^{IJ} + \epsilon^{IJ}\c^{0})\k^{J}, \;\;
\d_{\k}\Theta_{+}^{I} = \k_{+}^{I},\;\;
\d_{\k}\Theta_{-}^{I} = 0\,,
\ena
which shows that the gauge fixing for $\k$-symmetry should be
chosen as $\Theta_{+}^{I} = 0$ instead of $\Theta_{-}^{I} = 0$ 
in \p{gf}, that is, the usual Killing spinor
gauge is imcompatiable with the classical static vacuum solution
\p{ss} which was first mentioned in~\cite{PST}.

However, from \p{gf1} we know that in AdS coordinates \p{f2} the proper
gauge fixing for $\k$-symmetry is ${\cal P}_{+}\Theta = \Theta_{+} =
0$. Hence  we would like to see whether  choosing a different coordinate
frame could avoid the above imcompatibility. In the AdS coordinates
\p{wz} yields
\bea
I_{WZ} = \int dt \left(-\frac{Ar\dot{\tau}}{M} + BM\cos\chi \dot{\phi}
+ \frac{2Ar\bar\theta_{-}\c^{0}\theta_{-}}{M}\right),
\ena
where $\theta_{-}$ denotes $\theta_{-}^{1}$, and the $\k$-symmetry
gauge fixed super 0-brane action 
%% on AdS$_2 \times S^2$ in AdS coordinates is
becomes 
\bea
I_{0-brane} &=& \int dt \left\{-m\left[\left(\frac{r}{M}\right)^{2}
\left(\dot{\tau} + 2\bar\theta_{-}\c^{0}\dot{\theta}_{-}\right)^{2}
- \frac{M^{2}\dot{r}^{2}}{r^{2}} - M^{2}\left(\dot{\chi}^{2} +
\sin^{2}\chi \dot{\phi}^{2}\right)\right]^{\frac{1}{2}}\right\} \nn
&-& \int dt \left\{\frac{Ar}{M}\left(\dot{\tau} -
2\bar\theta_{-}\c^{0}\theta_{-}\right) - BM\cos\chi \dot{\phi}\right\}.
\label{ab}
\ena
When $B = 0$, to make the action \p{ab} vanish in the static solution
\p{ss}, we should choose $A = -m$. Then the $\k$-symmetry projector
is reduced to $\Gamma = {\cal E}\c^{0}$ and \p{ks} turns into
\bea
\d_{\k}\Theta^{I} = (\d^{IJ} - \epsilon^{IJ}\c^{0})\k^{J}, \;\;
\d_{\k}\Theta_{-}^{I} = \k_{-}^{I},\;\;
\d_{\k}\Theta_{+}^{I} = 0	\,,
\ena
which indicates that the gauge fixing for $\k$-symmetry should
be $\Theta_{-}^{I} = 0$ instead of $\Theta_{+}^{I} = 0$. Thus
the imcompatibility between the Killing gauge and the static
solution cannot be smoothed out by a change of the coordinate
frame. When the magnetic charge B is nonzero, however, this
imcompatibility can be avoided and the above simplification of
the super 0-brane action on AdS$_2 \times S^2$ in the Killing gauge
works, since in this case the static vacuum solution does
not exist.

The invertibility of the fermionic kinetic operator in action
\p{eb} can be seen from the quadratic term in the fermionic part
\bea
{\cal L}_{2} \sim \bar\theta_{+}\Omega\c^{0}\theta_{+}, \;\;
\Omega =
\left(\frac{2M}{r}\right)^{2}\left[e^{-1}\left(\frac{2M}{r}\right)
^{2}\dot{\tau} + A\right].
\ena
Since A is positive in the coordinate frame \p{f1}
with the static gauge $\tau = t$, $\Omega^{2}$ is nonzero, provided that
we choose the gauge fixing for e properly. 
This shows that the Killing gauge fixing is acceptable in the above derivation.

Up to now, we have obtained the $\k$-symmetry gauge fixed super 0-brane
action on AdS$_2 \times S^2$ with underlying $SU(1,1 \mid 2)$
invariance, which is the supersymmetric generalization of
the action in~\cite{CDK}. If we put  $\theta_{+} = 0$ and
$B = 0$, our bosonic Lagrangian is reduced to that in~\cite{CDK}
with $A = q$. Due to the restriction from $\k$-symmetry
\p{ma}, we have  $m = q$, but the other two cases for $m > q$ and $m < q$
cannot appear from our construction. It seems that we can add
the following term to the WZ term in order that we could introduce
more free parameters:
\bea
{\cal L}_{WZ} \rightarrow {\cal L}_{WZ} + \tilde{\cal H}
\ena
with
\bea
\tilde{\cal H}=  \tilde{A}\d^{IJ}L^{\a\alpr
I}(C\c)_{\a\b}C_{\alpr\bepr}\wedge L^{\b\bepr J}+\tilde{B}s^{IJ}L^{\a\alpr
I}C_{\a\b}(C\gapr)_{\alpr\bepr}\wedge L^{\b\bepr J},
\ena
where $s^{IJ}=(1, -1)$. One can easily check
\bea
d\tilde{\cal H}&=&0,\nn 
\d{\tilde{\cal H}}& = & -2d [{\tilde{A}}\d^{IJ}\bar{L}^{I}_{0}
(\c\otimes{1})\d{{\t}^J} + \tilde{B}s^{IJ}\bar{L}^{I}_{0}
({1}\otimes\gapr)\d{\t}^{J}],
\ena
so we can define a new $\k$-symmetry projector. However, when
we demand  $\C^{2} = 1$, we have  $\tilde{A} = \tilde{B} = 0$,
that is, we cannot introduce more free parameters to the action 
in this way. Then it is interesting to see whether it is possible
to find a more general super 0-brane action on AdS$_2 \times S^2$.

In~\cite{AS} it was observed that the quantum gravity on 
AdS$_2$ is a conformal theory on a strip which exhibits
the symmetries of the Virasoro algebra. Motivated by~\cite{AS},
the model in~\cite{CDK} was studied to see if one can find
the generators of the full Virasoro algebra~\cite{JK}. It was
shown that for the model in~\cite{CDK} with one dynamical
variable, one can find generators of the full Virasoro algebra,
but the central charge always vanishes. It would be interesting
to know whether there is a way to determine the central
charge and normal-ordering conventions for the present model, and to
see whether a non-vanishing central charge could be found
for the gauge-fixed super 0-brane model on  AdS$_2 \times S^2$,
which is related to the AdS$_2$/CF$T_1$ correspondence~\cite{AS}.

\section*{Acknowledgement} 

We would like to thank R. Manvelyan, N. Ohta and D. Sorokin for valuable 
discussions. 
This work is supported in part by the {\it Austrian 
Research Funds} FWF under grant Nr. M535-TPH.

%\newpage
\newcommand{\NP}[1]{Nucl.\ Phys.\ {\bf #1}}
\newcommand{\AP}[1]{Ann.\ Phys.\ {\bf #1}}
\newcommand{\PL}[1]{Phys.\ Lett.\ {\bf #1}}
\newcommand{\CQG}[1]{Class. Quant. Gravity {\bf #1}}
\newcommand{\CMP}[1]{Comm.\ Math.\ Phys.\ {\bf #1}}
\newcommand{\PR}[1]{Phys.\ Rev.\ {\bf #1}}
\newcommand{\PRL}[1]{Phys.\ Rev.\ Lett.\ {\bf #1}}
\newcommand{\PRE}[1]{Phys.\ Rep.\ {\bf #1}}
\newcommand{\PTP}[1]{Prog.\ Theor.\ Phys.\ {\bf #1}}
\newcommand{\PTPS}[1]{Prog.\ Theor.\ Phys.\ Suppl.\ {\bf #1}}
\newcommand{\MPL}[1]{Mod.\ Phys.\ Lett.\ {\bf #1}}
\newcommand{\IJMP}[1]{Int.\ Jour.\ Mod.\ Phys.\ {\bf #1}}
\newcommand{\JHEP}[1]{J.\ High\ Energy\ Phys.\ {\bf #1}}
\newcommand{\JP}[1]{Jour.\ Phys.\ {\bf #1}}

\end{document}